\documentclass[twocolumn,showpacs,preprintnumbers,amsmath,amssymb]{revtex4}
\usepackage{mathrsfs}
\usepackage{graphicx}
\usepackage{dcolumn}
\usepackage{bm}

\usepackage{color}

\begin{document}

\title{Influence of the Coulomb potential on above-threshold ionization: a quantum-orbit analysis beyond the strong-field approximation }

\author{X.-Y. Lai$^{1,2}$, C. Poli$^3$, H. Schomerus$^3$ and C. Figueira de Morisson Faria$^1$}
\affiliation{$^{1}$Department of Physics and Astronomy, University
College London, Gower Street, London WC1E 6BT, United Kingdom\\$^{2}$State Key
Laboratory of Magnetic Resonance and Atomic and Molecular Physics,
Wuhan Institute of Physics and Mathematics, Chinese Academy of
Sciences, Wuhan 430071, China\\$^3$ Department of Physics, Lancaster University, LA1 4YB Lancaster, United Kingdom }
\date{\today}

\begin{abstract}
We perform a detailed analysis of how the interplay between the residual binding potential and a strong laser field influences above-threshold ionization (ATI), employing a semi-analytical, Coulomb-corrected strong-field approximation (SFA) in which the Coulomb potential is incorporated in the electron propagation in the continuum. We find that the Coulomb interaction lifts the degeneracy of some SFA trajectories, and we identify a set of orbits which, for high enough photoelectron energies, may be associated with rescattering. Furthermore, by performing a direct comparison with the standard SFA, we show that several features in the ATI spectra can be traced back to the influence of the Coulomb potential on different electron trajectories. These features include a decrease in the contrast, a shift towards lower energies in the interference substructure, and an overall increase in the photoelectron yield. All features encountered exhibit a very good agreement with the \emph{ab initio} solution of the time-dependent Schr\"odinger equation.
\end{abstract}

 \maketitle

\section{Introduction}

When matter interacts with a strong laser field of peak intensity
around $10^{14}$ W/cm$^2$, the outmost electron may be freed by
absorbing many more photons than necessary. This very highly
nonlinear process is known as above-threshold ionization (ATI) and
has attracted considerable attention since the early work of Agostini and
co-workers \cite{Agostini1979PRL}; for a review see Ref.
\cite{Becker2002AdvAtMolOptPhys}. For typical parameters employed in experiments, i.e., near infra-red laser fields, it is commonly accepted that the electron reaches the continuum by tunnel ionization.
 If  the released electron revisits
the parent ion in the presence of the laser field
\cite{Schafer1993PRL,Corkum1993PRL}, this results in various
additional highly nonlinear phenomena, such as high-order ATI (HATI)
\cite{Paulus1994PRL}, high-order harmonic generation (HHG)
\cite{Ferray1988JPB}, and nonsequential double ionization (NSDI)
\cite{Walker1994PRL}.  Recently, considerable progress has been made
in the study of these nonlinear strong-field phenomena. For example,
both ATI and HHG have been employed as an important technique to
explore the electron shell structure and sub-femtosecond dynamics
\cite{Meckel2008Science,Itatani2004Nature,Kang2010PRL,Haessler2010NatP,Baker2006Science},
and NDSI has opened the door to the study of strong-field electron-electron
correlation \cite{Faria2011JMO,Becker2012RMP,Faria2012PRA,Hao2014PRL}.

In order to uncover the underlying physics of these highly nonlinear
phenomena, many theories and models have been proposed, such as
the \emph{ab initio} solution of the time-dependent Schr\"{o}dinger equation (TDSE) \cite{TDSE} and the quantum-orbit
theory within the strong-field approximation (SFA) \cite{LSH95,Carla2002PRA,Kopold2000OC}. Since the TDSE contains no physical approximation, its outcome is widely taken as a benchmark to evaluate the data in experiments and the
calculations of other theories and models \cite{Lindner2005PRL,Xie2012PRL,Bian2014PRL}. However, in many occasions the TDSE does not provide a transparent physical picture. Furthermore, since the numerical effort involved in \emph{ab initio} computations increases exponentially with the degrees of freedom, its implementation is impractical for strongly correlated multielectron systems. In contrast, the quantum-orbit theory provides very clear physical insight in terms of distinct electron ionization trajectories, and its outcome is qualitatively consistent with the experimental data. Therefore, it has been
widely and successfully used in the modeling of strong-field
phenomena \cite{LSH95,Carla2004PRA,Lai2013PRA,busuladzic2008PRA}

One should note, however, that the validity of the conventional quantum-orbit theory is limited. In fact, the use of the SFA before the application of the saddle-point approximation  implies that a considerable amount of physics is left out for the sake of a clear and intuitive picture \cite{Becker2002AdvAtMolOptPhys}. In particular,
the SFA fully neglects the effect of the Coulomb
potential of the parent ion on the ionized electrons,
approximating the continuum states by field-dressed plane waves \cite{Keldysh}. For single charged negative ions, this approximation is justified as the Coulomb interaction between the neutral core and the freed electron is absent. However, for atoms and molecules, this interaction is present, so that the SFA only works qualitatively. Furthermore, in recent years, several features have been observed which clearly highlight the influence of the Coulomb potential. Examples are the so-called low-energy structure (LES) in ATI spectra \cite{Moshammer2003PRL,Quan2009PRL,Blaga}, fan-shaped structures in photoelectron momentum distributions \cite{Rudenko2004JPB,Chen2006PRA,Arbo2006PRL,Arbo2008PRA,Yan2012PRA}, and the violation of the fourfold symmetry in angular electron distributions for elliptically polarized fields \cite{Bashkansky1988PRL,Popruzhenko2008PRA}.

Motivated by these observations, many methods have
been developed in the past few years in order to account for the Coulomb potential in orbit-based methods.  These include (i) using Coulomb-Volkov functions to describe the electron continuum states in the SFA \cite{Duchateau2002PRA,Milosevic1998PRA,arbo2014JPB}; (ii) incorporating the binding potential in the electron propagation using the eikonal Volkov approximation \cite{Smirnova2006JPB,Smirnova2008PRA}; (iii) a Coulomb-corrected SFA (CCSFA), which takes the trajectories from the SFA theory as a zero-order approximation and accounts for the Coulomb field perturbatively \cite{Popruzhenko2008JMO,Yan2010PRL,Yan2012PRA}; (iv) a quantum-trajectory Monte Carlo (QTMC),
which is based on Classical-Trajectory Monte-Carlo (CTMC) simulations, but considers the phase of each
trajectory \cite{Li2012PRL}; and (v) initial-value representations such as the Herman-Kluk propagator \cite{vandeSand1999PRL,Zagoya2012PRA,Zagoya2012NJP,Zagoya2014NJP} and the Coupled Coherent States method \cite{Guo2010PRA,Kirrander2011PRA,Symonds2015PRA}.

Most of the above-mentioned approaches have been applied to and
tested on direct ATI. This phenomenon is a particularly good testing
ground for Coulomb corrections for two main reasons. First, the
momentum range involved is relatively low, so that the influence of the
Coulomb potential is expected to be significant. Second, in contrast
to high-order ATI,  hard collisions are expected to be absent, so
that the Coulomb corrections are in principle easier to implement.
In particular the influence of the Coulomb potential on
quantum-interference patterns has attracted a great deal of
attention \cite{Yan2012PRA}. Furthermore, it has been shown that the
presence of the Coulomb potential considerably alters the topology
of the orbits, giving rise to types of trajectories that are absent
in the SFA \cite{Yan2010PRL}. In particular, in Ref.~\cite{Yan2012PRA} it
has been shown that sub-barrier corrections are necessary in order
to obtain the correct phases in the ATI electron momentum
distributions.

In this work, we develop a quantum-orbit theory with Coulomb
interactions  that, besides the effect on the phase, also accounts for the influence of the
interactions on the semiclassical amplitudes. We find that the amplitude is significantly modified both via the atomic
dipole moment at ionization and due to the altered stability of the trajectories during the electron propagation through the continuum.
 This Coulomb-corrected method is then employed to study the influence of the Coulomb potential on the
direct ATI ionization spectrum of Hydrogen. We perform a systematic investigation of how the Coulomb coupling changes the topology of the trajectories,  which are either decelerated or accelerated with regard to their SFA counterparts. This leads to a decrease in the phase difference between the contributions from different types of trajectories, which influences the interference patterns in the spectra. We also discuss how momentum non-conservation 
lifts the degeneracy of certain SFA trajectories. Furthermore, we verify that the distinction between direct and rescattered trajectories is blurred by the presence of the binding potential, which causes a set of trajectories to go around the core.

Our results show
that the spectrum calculated with this method is in much better agreement with the \emph{ab initio}
TDSE result than the predictions of the standard SFA.
In particular, the Coulomb-corrected theory recovers the
much weaker contrast in the interference substructure observed in the TDSE, and relates
this effect to the unequal semiclassical
weights of the electron trajectories in the presence of the Coulomb interactions.
Similarly to
what has been encountered in \cite{Yan2012PRA}, we also observe that the
positions of the interference maxima in the spectrum from the
quantum-orbit theory and TDSE result are shifted with respect to
the SFA simulations. However, our model indicates that these
shifts mainly stem from the modified electron propagation in the continuum.

This article is organized as follows. In Sec.~\ref{theory} we
describe the theoretical models employed in this paper, namely the
standard SFA and the Coulomb-corrected SFA, starting from the TDSE.
 In Sec.~\ref{results} we apply the theories to direct ATI and discuss the consequences of the Coulomb
interactions, first in terms of the individual trajectories and then
for the resulting ionization spectrum. Finally, in
Sec.~\ref{conclusions}, we summarize the main conclusions to be
drawn from this work. We use atomic units throughout.

\section{Theoretical models}
\label{theory}
The underlying framework for the subsequent discussions is the time-dependent Schr\"odinger equation
\begin{equation}
i\partial_t|\psi(t)\rangle=H(t)|\psi(t)\rangle \,.
\end{equation}
In the ionization problems considered in this work, the Hamiltonian separates into two parts, $H(t)=H_a+H_I(t)$. Here
\begin{equation}
H_a=\frac{\hat{\mathbf{p}}^{2}}{2}+V(\hat{\mathbf{r}})
\end{equation}
denotes the field-free one-electron atomic Hamiltonian and the hats denote operators. In the problem addressed here, we consider a 
Coulomb-type potential
\begin{equation}
V(\hat{\mathbf{r}})=-\frac{C}{\sqrt{\hat{\mathbf{r}}\cdot \hat{\mathbf{r}}}},\label{eq:potential}
\end{equation}
where 
$0\leq C\leq 1$ is an effective coupling, which we vary in a continuous fashion in order  to assess the influence of the Coulomb potential. For  Hydrogen, $C=1$.  Furthermore, $H_I(t)$ describes the interaction with the laser field. In the velocity and length gauges, this interaction is given by $H_I(t)=\hat{\mathbf{p}}\cdot \mathbf{A}(t)+\mathbf{A}^2/2$ and $H_I(t)=-\hat{\mathbf{r}}\cdot \mathbf{E}(t)$, respectively, where $\mathbf{E}(t)=-d\mathbf{A}(t)/dt $
is the external laser field. The length gauge provides us with the physical picture of ionization as a tunneling process driven by an effective time-dependent potential. This gauge will be employed throughout.

The time-evolution operator associated with this Hamiltonian is of the general form
\begin{equation}
U(t,t_0)=\mathcal{T}\exp \bigg [i \int^t_{t_0}H(t^{\prime})dt^{\prime} \bigg],
\end{equation}
where $\mathcal{T}$ denotes time-ordering. This operator takes a wave function from a time $t_0$ to a time $t$, i.e., $|\psi(t)\rangle=U(t,t_0)|\psi(t_0)\rangle$, and satisfies
\begin{eqnarray}
i \partial_t U(t,t_0) &= &H(t) U(t,t_0) \, ,\nonumber\\
-i \partial_{t_0} U(t,t_0) &= &U(t,t_0)H(t_0)\, .
\end{eqnarray}
Employing the Dyson equation, the time-evolution operator may be written as
\begin{equation}
U(t,t_0)=U_a(t,t_0)-i\int\displaylimits^t_{t_0}U(t,t^{\prime})H_I(t^{\prime})U_a(t^{\prime},t_0)dt^{\prime}\,
,
\end{equation}
where $U_a(t,t_0)$ is the time-evolution operator associated with the field-free Hamiltonian.

For above-threshold ionization, the initial state is a bound state $\left\vert \psi _{0}\right\rangle $, while the final state is a continuum state $ |\psi_{\textbf{p}}(t)\rangle$ with drift
momentum $\mathbf{p}$. This gives the ionization amplitude \cite{Becker2002AdvAtMolOptPhys}
\begin{equation}\label{Mpdir}
M(\mathbf{p})=-i \lim_{t\rightarrow \infty} \int\displaylimits_{-\infty }^{t }d
t^{\prime}\left\langle \psi_{\textbf{p}}(t) |U(t,t^{\prime})H_I(t^{\prime})| \psi
_0(t^{\prime})\right\rangle \, ,
\end{equation}
which is formally exact.

\subsection{Strong-field approximation}
Equation ({\ref{Mpdir}}) cannot be solved in closed form, so that approximations are required in order to compute the ATI transition amplitude via analytical methods. A  popular approximation is to replace $U(t,t^{\prime})$ by the Volkov time evolution operator $U^{(V)}(t,t^{\prime})$ in Eq.~({\ref{Mpdir}}). This  implies that the continuum has been approximated by Volkov states, i.e., by field-dressed plane waves $|\psi^{(V)}_{\textbf{p}}(t) \rangle$, where
\begin{equation}
\langle\mathbf{r}|\psi^{(V)}_{\textbf{p}}(t)\rangle=\langle\mathbf{r}|\mathbf{p}(t)\rangle\exp\left[-i\int^t_{-\infty} d\tau\frac{[\mathbf{p}+\mathbf{A}(\tau)]^2}{2}\right]
\label{Volkov}
 \end{equation}
with
 \begin{equation}
 \langle\mathbf{r}|\mathbf{p}(t)\rangle=\frac{\exp[i \mathbf{p}(t)\cdot \mathbf{r}]}{(2 \pi)^{3/2}}.
\end{equation}
Here $\mathbf{p}(t)=\mathbf{p}+\mathbf{A}(t)$ in the length gauge and  $\mathbf{p}(t)=\mathbf{p}$ in the velocity gauge, so that $U^{(V)}(t,t^{\prime})|\psi_{\textbf{p}}(t^{\prime}) \rangle=|\psi^{(V)}_{\textbf{p}}(t) \rangle$ \footnote{The phase $\exp[-i \mathbf{A}(t)\cdot \mathbf{r}]$ associated with the gauge transformation will cancel out that existent in the Volkov state. }. This is the key idea behind the strong-field approximation or Keldysh-Faisal-Reiss theory \cite{Keldysh,Faisal1973JPB,Reiss1980PRA}.
For detailed discussions see, e.g., \cite{Becker1997PRA,Fring1996JPB} and the  recent tutorials \cite{Smirnova2007JMO}. Within the SFA, the amplitude ({\ref{Mpdir}}) is then given by \cite{Keldysh,Faisal1973JPB,Reiss1980PRA,Becker2002AdvAtMolOptPhys}
\begin{equation}\label{MpSFA}
M(\mathbf{p})=-i\int\displaylimits_{-\infty }^{\infty }dt^{\prime}\left\langle
\mathbf{p}+\mathbf{A}(t^{\prime})\right. |H_I(t^{\prime})|\left. \psi
_{0}\right\rangle e^{ iS(\mathbf{p},t^{\prime})}.
\end{equation}
Here
\begin{equation}
\label{action}
S(\mathbf{p},t^{\prime})=-\frac{1}{2}\int_{t^{\prime}}^{\infty }[\mathbf{p}+
\mathbf{A}(\tau )]^{2}d\tau +I_{p}t^{\prime}
\end{equation}
is the semiclassical action, where $I_{p}$ gives the ionization potential
and $\mathbf{A}(t)$ denotes the vector potential of the laser field. In Eq.~(\ref{MpSFA}), we have also employed the notation
$|\psi_0(t^{\prime})\rangle=e^{iI_pt^{\prime}}|\psi_0\rangle$.

For sufficiently high intensity and low frequency of the laser
field, the temporal integration in Eq.~(\ref{MpSFA}) can be
evaluated by the saddle-point method \cite{Carla2002PRA,Kopold2000OC}, which seeks solutions
such that the action (\ref{action}) is stationary. The corresponding saddle-point
equation reads
\begin{equation} \label{tun_time_sfa}
\frac{[ \textbf{p}+\textbf{A}(t^{\prime})]^{2}}{2}+I_{p}=0.
\end{equation}
Physically, Eq.~(\ref{tun_time_sfa}) ensures the conservation of energy at
the ionization time $t^{\prime}$, which leads to complex solutions $t_s$.
In terms of these solutions,
the transition amplitude (\ref{MpSFA}) can then be written as
\begin{equation} \label{sfa_saddle}
M(\mathbf{p}) \sim   \sum_{s}\mathcal{C}(t_s)\left\langle
\mathbf{p}+\mathbf{A}(t_s)\right. |H_I(t_s)|\left. \psi
_{0}\right\rangle e^{iS(\mathbf{p},t_s)} \, ,
\end{equation}
where the prefactors
\begin{equation}
\label{eq:sfaamp}
\mathcal{C}(t_s)=\sqrt{\frac{2\pi i}{\partial
^{2}S(\mathbf{p},t_{s})/\partial t_{s}^{2}}}
\end{equation}
are expected to vary much more slowly than the action for the saddle-point approximation to hold. Since each solution $t_s$ represents a distinct trajectory of the electron in the laser
field, the sum in Eq.~(\ref{sfa_saddle}) denotes the interference
between different quantum paths, which has been extensively studied
in the literature (for reviews see, e.g., \cite{Becker2002AdvAtMolOptPhys,SalieresScience2001}). One should note that in the SFA, the field-dressed momentum is conserved.

\subsection{Coulomb-corrected SFA}

Within the SFA, an electron no longer feels the atomic potential
after it has been promoted into the continuum at the time $t_s$,
resulting in the considerable deviations between this model and
experimental results. In this section we describe a
Coulomb-corrected SFA (CCSFA) which cures this shortcoming of the
SFA (for similar approaches see
\cite{Smirnova2006JPB,Smirnova2008PRA,Popruzhenko2008PRA,Popruzhenko2008JMO}).

First we note that in presence of the Coulomb potential, momentum is no longer conserved, and the time evolution operator depends on both $\hat{\mathbf{r}}$ and $\hat{\mathbf{p}}$. As a result, the time evolution operator cannot be diagonalized by the Volkov states (\ref{Volkov}).

Instead, adopting Feynman's path integral formalism \cite{Kleinert2009} the Coulomb corrected transition amplitude reads
\begin{multline}\label{1}
M(\mathbf{p})=-i\lim_{t\rightarrow \infty }\int_{-\infty }^{t}dt^{\prime} \int_{\mathbf{p}(t')}^{\mathbf{p}(t)}
 \mathcal {D} \textbf{p} \int \frac{\mathcal {D} \textbf{r}}{(2\pi)^2} \\
  \times  e^{i S(\textbf{p},\textbf{r},t^{\prime}, t)}
\left\langle \mathbf{p}_0+\mathbf{A}(t^{\prime})\right. |H_I (t^{\prime})|\left. \psi_{0}\right\rangle\,,
\end{multline}
where the action is given by
\begin{equation}\label{2}
S(\textbf{p},\textbf{r},t^{\prime}, t) =I_pt^{\prime}+\tilde{S}(\textbf{p},\textbf{r},t^{\prime}, t),
\end{equation}
with
\begin{equation}\label{stilde}
 \tilde{S}(\textbf{p},\textbf{r},t^{\prime}, t)=-\int^t_{t^{\prime}}[\mathbf{r}(\tau) \cdot \dot{\mathbf{p}}(\tau)+H(\mathbf{r}(\tau),\mathbf{p}(\tau),\tau)]d\tau,
\end{equation}
and
\begin{equation}
H(\mathbf{r}(\tau),\mathbf{p}(\tau),\tau)=\frac{1}{2}\left[\mathbf{p}(\tau)+\mathbf{A}(\tau)\right]^2 -\frac{C}{\sqrt{\mathbf{r}(\tau)\cdot\mathbf{r}(\tau)}}.
\label{Hamiltonianpath}
\end{equation}
One should note that the problem is solved in the length gauge, and
Eq.~(\ref{Hamiltonianpath}) can be obtained from the standard
length-gauge Hamiltonian by a partial integration. This issue has
been discussed in detail in Ref.~\cite{Popruzhenko2014JPB}.

Following the same procedure as for the SFA, we can now obtain the
Coulomb-corrected transition amplitude by applying the saddle-point
approximation.  By construction, the saddle-point equation on $t'$
leads to the condition
\begin{equation}
\label{tun_time_ccsfa}
\frac{[ \textbf{p}+\textbf{A}(t^{\prime})]^{2}}{2}+I_{p}+V[\mathbf{r}(t^{\prime})]=0.
\end{equation}
The Coulomb-corrected transition amplitude is then given by
\begin{multline}\label{3}
M(\mathbf{p})=-i\sum_{s}\sqrt{\frac{2\pi i}{S^{''}(t_s)}} \int_{\mathbf{p}(t_s)}^{\mathbf{p}(t)}
 \mathcal {D} \textbf{p} \int \frac{\mathcal {D}\textbf{r}}{(2\pi)^2} \\ \times  e^{i S(\textbf{r}, \textbf{p}, t_s, t)}
\left\langle \mathbf{p}(t_s)+\mathbf{A}(t_s)\right. |H_I (t_s)|\left. \psi
_{0}\right\rangle\,.
\end{multline}
We then use the semi-classical path integral theory developed by Van Vleck and Gutzwiller \cite{Kay2013} on the paths $[\mathbf{p}(t)]$, and $[\mathbf{r}(t)]$. The associated saddle-point equations take the form of classical equations of motion for the trajectories, \begin{equation}\label{saddle1}
\textbf{\.{p}}(\tau)= -\nabla_\textbf{r}V[\textbf{r}(\tau)]\, ,
\end{equation}
\begin{equation}\label{saddle2}
\textbf{\.r}(\tau)= \textbf{p}(\tau)+\textbf{A}(\tau)\, .
\end{equation}
The solutions of these equations in general are again complex due to the additional condition \eqref{tun_time_ccsfa}.

In terms of such solutions, the Coulomb-corrected transition amplitude finally reads
\begin{multline}
\label{MpPathSaddle}
M(\mathbf{p})=-i\sum_{s}\frac{1}{2\pi i\sqrt{S^{''}(t_s)}} \bigg\{\det \bigg[  \frac{\partial\mathbf{p}_s(t)}{\partial \mathbf{r}_s(t_s)} \bigg] \bigg\}^{-1/2} e^{i\nu}\\
  \times  e^{i S(\textbf{r}_s, \textbf{p}_s, t_s, t)} \left\langle \mathbf{p}_s(t_s)+\mathbf{A}(t_s)\right. |H_I (t_s)|\left. \psi
_{0}\right\rangle\,.
\end{multline}
The sum is over the classical trajectories that begin at position
$\mathbf{r}(t_s)$ at time $t_s$, and end at momentum $\mathbf{p}(t)$
at time $t\rightarrow \infty$ (sum over multiple solutions with
identical $t_s$ is implied). Each trajectory contributes a term with
phase given by the Maslov index  $\nu$ and the classical action
$S(\textbf{r}_s, \textbf{p}_s, t_s, t)$ given by Eq.~(\ref{2}) at
the stationary values $t_s$, $\mathbf{p}_s$, $r_s$. In practice, $t$
is defined at the end of the pulse, which should be taken to be
sufficiently long.

Due to the presence of the binding potential, the above-stated
equation exhibits branch cuts for
$\mathrm{Re}[\mathbf{r}_s(\tau)\cdot\mathbf{r}_s(\tau)]<0$ and
$\mathrm{Im}[\mathbf{r}_s(\tau)\cdot\mathbf{r}_s(\tau)]=0$.  For
vanishing transverse momenta, the branching points turn into
first-order poles and this problem is absent (for details on these
branch cuts see Ref.~\cite{Popruzhenko2014JETP}). These branch cuts
can be avoided if one takes the integration contour along the real
time axis once the electron is in the continuum. More specifically,
the time integration contour is taken first parallel to the
imaginary time axis, up to the so-called ``tunnel exit", i.e., the
point in space at which the electron tunnels out of the potential
barrier, and then along the real time axis. This is the procedure
taken by most groups when implementing Coulomb-corrected
strong-field theories (see, e.g.,
Refs.~\cite{Yan2012PRA,Torlina2013PRA}).

Furthermore, besides the SFA factor \eqref{eq:sfaamp} and the tunnel
matrix element $\left\langle
\mathbf{p}_s(t_s)+\mathbf{A}(t_s)\right. |H_I (t_s)|\left. \psi
_{0}\right\rangle$, the amplitude now involves the stability
$\frac{\partial\mathbf{p}_s(t)}{\partial \mathbf{r}_s(t_s)}$ of the
trajectories. In the limit of vanishing binding potential, the usual
SFA is recovered \cite{Milo2013JMP}. However, as we will see, this
happens in a nontrivial fashion when the degeneracy breaking of
trajectories is taken into account.

\section{Results and discussion}
\label{results}

In the results that follow, we use the monochromatic laser field
\begin{equation}
\mathbf{E}(t)= \text{\textbf{\^{z}}}    E_0 \sin \omega t.
\end{equation}
For this type of field, $\textbf{A}(t)\propto \cos(\omega
t)\textbf{\^{z}}$  and given final momentum $\mathbf{p}$, there are
two solutions $t_s$ of Eq.~(\ref {tun_time_sfa}) for the SFA per
cycle of the laser field. In \cite{Yan2010PRL}, these solutions have
been related to Orbits I and II, depending on whether the electron
leaves in the same or in the opposite direction to the detector. In
the results that follow, we will consider this classification and
its extension to the Coulomb-corrected case.

The initial states are taken as the ground state of Hydrogen, i.e.,
$\psi _{0}(r)= \langle \mathbf{r} | \psi _{0}\rangle=e^{-r}/ \sqrt{\pi}$. In this case, the tunnel matrix element in  Eq.~(\ref{MpPathSaddle}) becomes related to the atomic
dipole moment and can be simplified as $
\left\langle
\mathbf{p}_{s}(t_s)+\mathbf{A}(t_s)\right. |-\mathbf{r}\cdot \mathbf{E}%
(t_s)|\left. \psi _{0}\right\rangle\sim E(t)k_z$, where
$\textbf{k}=\mathbf{p}_{s}(t_s)+\mathbf{A}(t_s)$ \cite{Milosevic2006JPB}. Unless otherwise stated, we will consider $C=1$ in $V(r)$.

\subsection{Coulomb-corrected saddle-point solutions and their physical implications}
In comparison with the SFA, the canonical momentum $\textbf{p}$
of the electron is time-dependent according
Eq.~(\ref{saddle1}) if the Coulomb interaction is incorporated.  Therefore, in the CCSFA theory, the greatest challenge is to
solve the saddle-point equations for the tunneling time $t^{\prime}$ and the
canonical momentum $\textbf{p}_0$ for any  given final momentum $\textbf{p}$. One should also bear in mind that, in experiments, the measured photoelectron spectra is a function of the final momentum. Therefore, if, for a given final momentum,
the initial conditions for the corresponding electron trajectories could be obtained reversely, it would be easier to understand how these trajectories were influenced by the Coulomb potential.

In order to simplify the calculation and isolate the main effects of
the potential on the trajectories, we assume that the electron is
ionized by tunneling from the time $t_s$ to $t_s^R= \mathrm{Re}\,
t_s$ at a fixed momentum $p_s(t_s)$  and then moves to detector with
the real time and coordinate according to the classical
equations of motion  \eqref{saddle1} and  \eqref{saddle2}. This is
the most widely used assumption for the contour, and has been
employed in \cite{Popruzhenko2008JMO,Yan2010PRL,Yan2012PRA} (for a
review see \cite{Popruzhenko2014JPB}). Within this set of
assumptions, it is a reasonable approximation to neglect the Coulomb
potential in Eq.~\eqref{tun_time_ccsfa}, which thus reduces to
Eq.~\eqref{tun_time_sfa}. The tunneling exit at the time $t_R$ is
given by
\begin{equation}\label{exit}
z_0=\alpha(t_s^R)-\mathrm{Re}\,\alpha(t)\,,
\end{equation}
where $\alpha(t)=\int ^t\textbf{A}(\tau) d\tau$ \cite{PPT1967}.
\begin{figure}[tp]
\begin{center}
\includegraphics[scale=0.3]{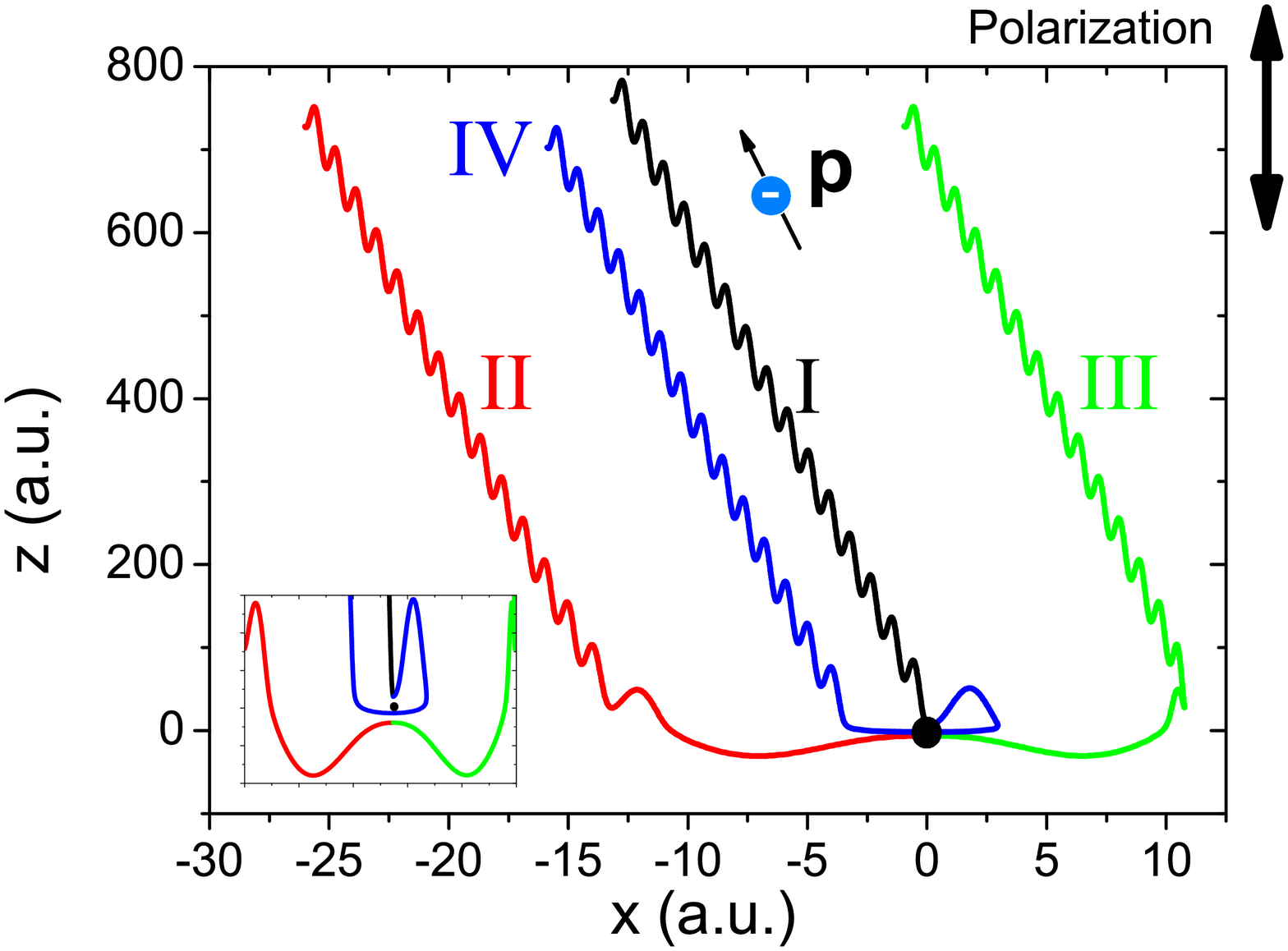}
\end{center}
\caption{(Color online) Illustration of the four types of CCSFA
trajectories in the $zx$ plane for electrons with fixed final
momentum $\textbf{p}$, computed using a linearly polarized square
pulse of duration $T_p=15.25$ cycles,  intensity $I=2 \times 10^{14}
\mathrm{W/cm}^2$ and frequency $\omega=0.057$ a.u., and a Coulomb
potential (\ref{eq:potential}) with $C=1$. The ionization potential
has been taken as $I_p=0.5$ a.u. The black dot denotes the position
of the nucleus. The inset shows the region near the core. }
\label{fig1}\end{figure}

Figure \ref{fig1} depicts four types of the trajectories in the $zx$ plane
for an electron with a given final momentum $\textbf{p}$.  For
trajectories of type I, the tunneling exit $z_0>0$, and the electron moves
directly towards the detector without returning to its parent
core. For the type II and III trajectories, the tunneling exit
$z_0<0$, meaning that the initial motion carries the electron away from the detector before it turns around and ends up with the stipulated momentum $\textbf{p}$.  A closer inspection shows that they are similar to Kepler hyperbolae to which a drift motion caused by the field is superimposed \cite{Yan2010PRL,Arbo2006PRL}.
Trajectory types I and II are similar to the so-called ``short'' and ``long''
trajectories in the SFA theory. The type III is not found in the  SFA
and can be observed after the Coulomb potential is considered, which
is consistent with earlier work \cite{Yan2010PRL}.

\begin{figure}[tb]
\begin{center}
\includegraphics[width=.8\columnwidth]{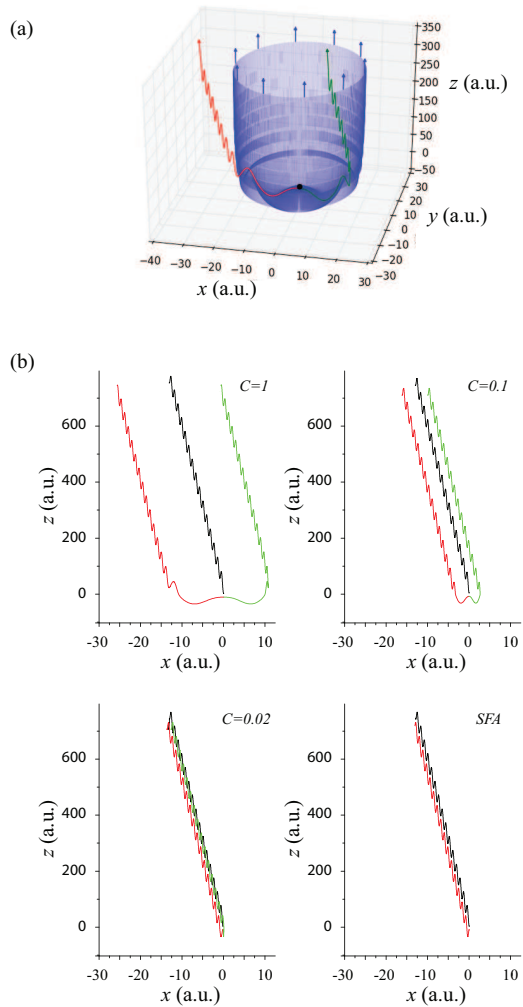}
\end{center}
\caption{(Color online) Emergence of the two distinct trajectories
II and III. (a) For ionization parallel to the field, the
trajectories are part of a torus, respecting the rotational
invariance of the system. In the SFA, this torus would contract to a
point as momentum conversion then dictates that the initial momentum
vanishes. At a finite tilt angle of the final momentum (here
$2^\circ$), however, the torus splits into two distinct solutions,
II and III (red and green). (b) As the Coulomb interaction strength
$C$ is then reduced, trajectories II and III merge, and remain
distinct from trajectory I (here the tilt angle of the final
momentum is $1^\circ$). The remaining parameters are the same as in
the previous figure.} \label{fig:charles} \label{fig2}\end{figure}

As we show in Fig.~\ref{fig:charles}(a), the emergence of the new
trajectory is directly related to the momentum non-conservation. It
is instructive to start with the case of ionization parallel to the
field, where the final momentum $\textbf{p}$ is along the
polarization direction. On first inspection, the type II and III
trajectories are symmetric with respective to the polarization
direction. As a matter of fact, trajectories II and III then
degenerate into a torus, with a finite initial transverse momentum.
In the SFA, the torus contracts onto a single trajectory with a
vanishing initial transverse momentum, as dictated by momentum
conservation. At a finite tilt angle, the torus splits in analogy to
the Poincar{\'e}-Birkhoff scenario in KAM theory, leaving two
clearly distinct trajectories. As we change the effective Coulomb
interaction strength $C$ from $C=1$ (Hydrogen) to $C=0$, the
situation from the SFA is again recovered [panel (b) in
Fig.~\ref{fig:charles}].

\begin{figure}[tb]
    \begin{center}
        \includegraphics[width=.8\columnwidth]{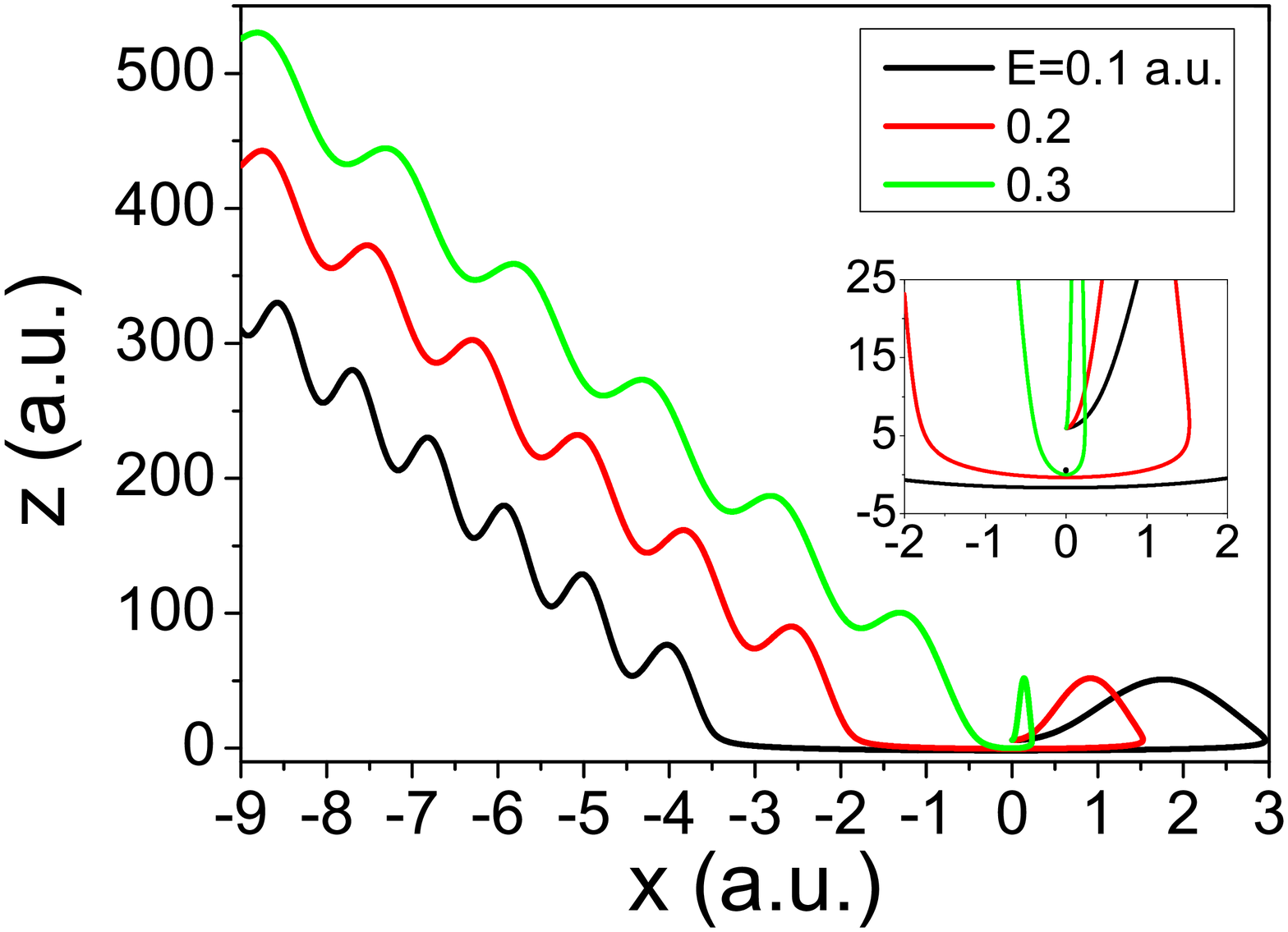}
    \end{center}
    \caption{(Color online) Behavior of type IV orbits for increasing photoelectron energy. The field and atomic parameters are the same as in the previous figures, and the energy is indicated on the upper right corner. A blow-up of the figure shows the trajectories approaching the origin, which is marked by a black dot. The field and atomic parameters are given in Fig.~\ref{fig1}.}
    \label{fig:fourthorbit}
\end{figure}

It  is noteworthy that our numerics uncover an additional trajectory
type, denoted as  IV. Although the tunnel exit points towards the
detector, the electron is driven back to the core by the laser
field, then goes around the core, and finally moves towards to the
detector. With increasing photoelectron final momentum $\mathbf{p}$,
the shortest distance between the electron and the core decreases.
This distance can be smaller than the tunnel exit. In this case,
this type of trajectory corresponds to a rescattering event. This
behavior is shown in Fig.~\ref{fig:fourthorbit}.

Since the contribution of events involving rescattering to the
low-energy photoelectron spectra is small, it can be safely
neglected, so that we only need to consider the type I-III
trajectories. However, it is encouraging to see that rescattering
contributions already show up within a framework that is formally
tailored for direct ATI, even though they eventually may require a
more accurate treatment including rescattering form factors  that
account for the inherently diffractive, nonclassical nature of these
events \cite{Carla2002PRA,Carla2004PRA}.

\subsection{Photoelectron spectrum}

Based on  the trajectories described in the previous subsection we
now study the photoelectron spectrum within the CCSFA theory and
compare the result with the standard SFA, while taking the \emph{ab
initio} TDSE calculation as a benchmark. The TDSE has been computed
using the freely available software Qprop \cite{Qprop}. The standard
SFA is implemented according to Eq.~(\ref{sfa_saddle}).  Within the
CCSFA, we calculate the stability of the trajectories numerically.
In practice, instead of using $\partial \mathbf{p}_s(t)/\partial
\mathbf{r}_s(t_s)$ in Eq.~(\ref{MpPathSaddle}) we employ $\partial
\mathbf{p}_s(t)/\partial \mathbf{p}_s(t_s)$. The latter stability
factor is of easier implementation, and can be obtained using a
Legendre transformation in the transition amplitude (\ref{3}). Upon
this transformation, the action will remain the same as long as the
electron starts from the origin, which is one of the assumptions
considered in this work. For details on Legendre transformations
see, e.g., \cite{Goldstein}.

We use the tunnel exit approximation \eqref{exit} to split the
action into a part inside the barrier,
\begin{equation}\label{in_barrier}
\tilde{S}_s^{\text{in}}(\textbf{p}_s,\textbf{r}_s,t_s)=-\int_{t_s}^{t_s^R} H(\textbf{p}_s,\textbf{r}_s,\tau)d \tau
\end{equation}
with the
 tunneling trajectory
 $\textbf{r}_s(t)=\int_{t_s}^{t}
[\textbf{p}_{s}(\tau)+\textbf{A}(\tau)]d \tau $
\cite{Popruzhenko2008JMO}, and a part outside the barrier,
\begin{equation}
\tilde{S}_s^{\text{out}}(\textbf{p}_s,\textbf{r}_s,t_s)=\int_{t_s^{R}}^{T_p}d
\tau [-\dot{\textbf{p}}_s(\tau)\cdot
\textbf{r}_s(\tau)-H(\textbf{p}_s,\textbf{r}_s,\tau)],
\end{equation}
with the ionization trajectory determined as described in the
previous subsection. Note that, in Eq.~(\ref{in_barrier}), the term
in $\dot{\mathbf{p}}_s$ is vanishing as the momentum inside the
barrier was taken to be constant. Furthermore, real variables
outside the barrier will lead to real stability factors, so that
phase differences in the continuum stem exclusively from the action.

\begin{figure}[tp]
\begin{center}
\includegraphics[scale=0.3]{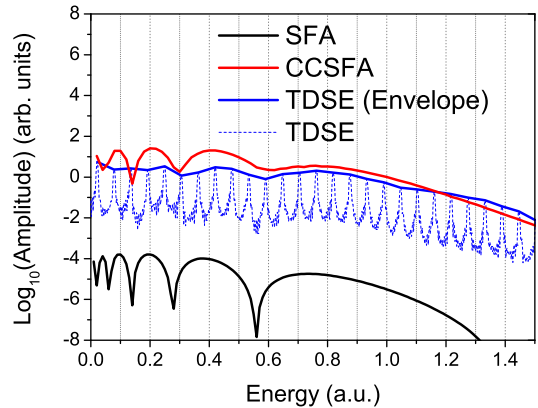}
\end{center}
\caption{(Color online) Photoelectron spectra for ionization along the laser polarization direction, with
the standard SFA, the Coulomb-corrected CCSFA, and from \emph{ab initio}
TDSE calculation. The blue solid curve is the envelope of the TDSE
calculation. In order to perform a clearer comparison, the TDSE spectrum has been shifted upwards in around one order of magnitude. The laser-field and atomic  parameters are given in Fig.~\ref{fig1}.}\label{fig2}
\end{figure}

The resulting photoelectron spectra in the direction along the laser polarization are shown in Fig.~\ref{fig2}.
Since the SFA and CCSFA only accounts for the
interference of the electrons ionized in one optical cycle, the spectra correspond to an envelope, without the
sharp ATI peaks seed in the \emph{ab initio}
calculation. We therefore also show the envelope of the spectrum from the
\emph{ab initio} method. Clear
interference structures are observed in all spectra. A closer
inspection shows that the interference contrast in the spectra from the TDSE and
the CCSFA theory is much weaker than that from the SFA theory. Moreover, the
positions of the interference maxima in the CCSFA spectrum are in a better
agreement with the TDSE result than the SFA. 

\begin{figure}[tb]
\begin{center}
\includegraphics[scale=0.3]{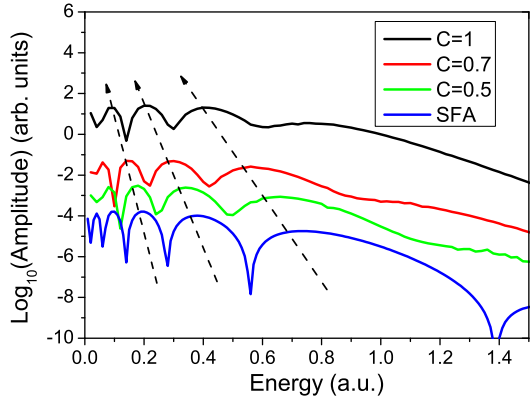}
\end{center}
\caption{ (Color online) Photoelectron spectra from the CCSFA theory with different
strengths $C$ of the rescaled Coulomb potential given by Eq.~(\ref{eq:potential}). 
 For $C=0$ the spectrum coincides with that of the SFA theory. The remaining laser-field and atomic  parameters are given in Fig.~\ref{fig1}.}\label{fig6}
\end{figure}

The  mechanisms leading to these improvements are
explained in the next subsections. For reference, it is useful to inspect how the Coulomb corrections are established when one changes the effective interaction strength $C$, so that for $C=0$ the SFA is recovered. Fig.~\ref{fig6}
shows the photoelectron spectra for different values of $C$.

\subsubsection{Interference contrast}

\begin{figure}[tb]
\begin{center}
\includegraphics[scale=0.3]{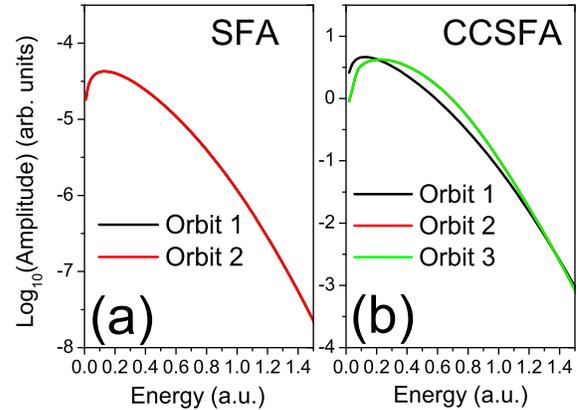}
\end{center}
\caption{ (Color online) Amplitude of each trajectory type as a function of the
photoelectron energy (a) within the SFA theory and  (b) within the  CCSFA
theory. The laser-field and atomic  parameters are given in Fig.~\ref{fig1}.}\label{fig3}
\end{figure}

First, we consider the effect of the Coulomb potential  on  the interference contrast. According to the
discussion above, the type I-III trajectories are dominant for the
electrons with low kinetic energy. For the photoelectron spectrum along the laser
polarization, type II and III trajectories are symmetric with
respect to the polarization direction and have the same phase and
amplitude. Therefore, the interference pattern in the
spectrum arises from the beating between type I trajectory and
type II and III trajectories. In Fig.~\ref{fig3}, we present the
amplitude related to each orbit as a function of the photoelectron energy
with and without Coulomb corrections, respectively.

In the SFA theory, the amplitudes associated with the type I and the
type II trajectories are the same. This holds because in the SFA the
electron's final momentum is solely determined by the ionization
time, which for trajectories I and II are displaced by half a cycle.
This means that the absolute values of the electric field, and hence
the ionization probability, are the same. Therefore, in the SFA the
interference contrast will be maximal. If the Coulomb potential is
included, the amplitudes of the type I and II/III trajectories
differ slightly. Furthermore, the
joint amplitude of the type II and III trajectories exceeds that of
type I significantly. All this leads to a much reduced contrast of
the interference pattern \footnote{In principle, the
Poincar{\'e}-Birkhoff type scenario has implications for the
saddle-point treatment \cite{Tomsovic,Ullmo}, in analogy to the
uniform approximations needed close to orbit bifurcation scenarios
\cite{Carla2002PRA,Carla2004PRA}, but as we do not encounter any
divergence and in view of providing a simple picture, we did not
find it necessary to employ this here.}.

\begin{figure}[tb]
\begin{center}
\includegraphics[scale=0.32]{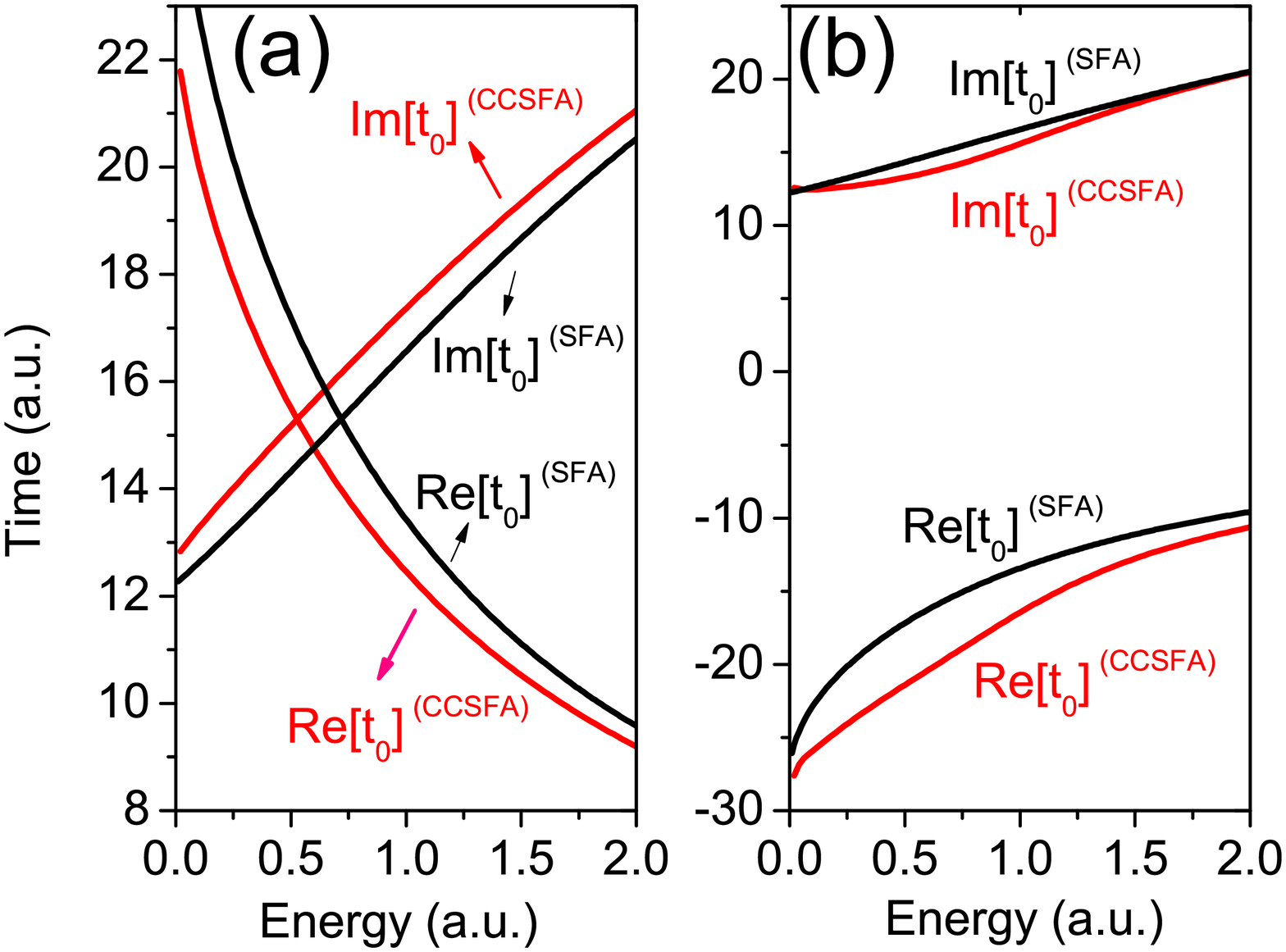}
\end{center}
\caption{ (Color online) Time of tunneling as a function of the photoelectron
energy. (a) for the type I orbit and (b) for the type II orbit. The
time is separated into two parts: real part and imaginary part. The
black curves are for the SFA calculation and the red curves are from the
CCSFA. The laser-field and atomic  parameters are given in Fig.~\ref{fig1}.}\label{fig}
\end{figure}

\begin{figure}[tb]
\begin{center}
\includegraphics[scale=0.3]{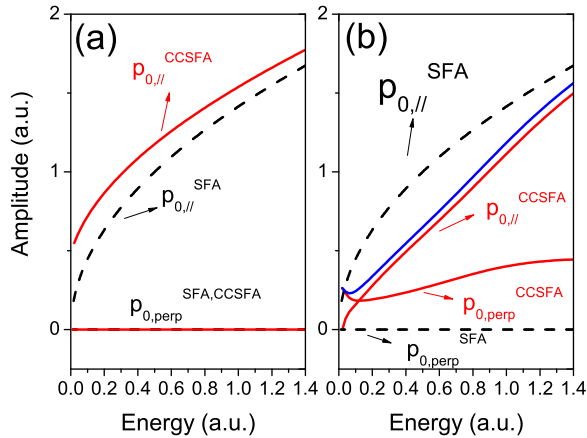}
\end{center}
\caption{(Color online) Initial momentum for trajectories of type I (a) and II (b) as a function of
the photoelectron energy. The momentum is separated into the parallel and perpendicular components.  The blue
curve in (b) denotes the total momentum from the CCSFA theory. The laser-field and atomic  parameters are given in Fig.~\ref{fig1}.}\label{fig4}
\end{figure}

Phenomenological insight into the amplitude difference between the
type I and type II/III trajectories can be obtained by considering
the  tunneling time and initial momentum  as a function of the
photoelectron energy. The imaginary part $\mathrm{Im}\,[t_s]$ can be
interpreted as the time it takes the electron to tunnel through the
potential barrier \cite{hauge}. The larger $\mathrm{Im}\,[t_s]$ is,
the lower the ionization rate. Fig.~\ref{fig} exhibits the time of
tunneling as a function of the photoelectron energy for the type I
orbit and the type II orbit, respectively. For the type I orbit,
$\mathrm{Im}\,[t_s]$ increases when the Coulomb potential is taken
into account. In contrast, for
type II and III trajectories, Im$[t_s]$ is smaller in the CCSFA than
in the SFA. Therefore, if the Coulomb corrections are present, the
amplitude of the type I will be smaller than that of type II. These features are consistent
with the changes in $\mathrm{Re}[t_s]$, which, for Orbit I, move towards the field crossing and, for Orbits II/III, is displaced towards the times for which the field amplitude is maximal. This implies that the effective potential barrier will be wider for the former orbits and narrower for the latter.

According to the Ammosov-Delone-Krainov (ADK) theory \cite{ADK},
these observations can be linked to the initial momentum, with a
large momentum translating into a reduced  ionization rate.
Fig.~\ref{fig4}(a) shows that for trajectories of type I, the
initial momentum from CCSFA theory is indeed larger than that from
the SFA calculation. This can be understood from the fact that the
electron needs to compensate the deceleration in the Coulomb
potential as it moves towards the detector. In contrast, for type II
and III trajectories, the initial parallel momentum from the CCSFA
theory is smaller than in  the SFA. Although there is a nonvanishing
perpendicular momentum from the CCSFA theory, the total initial
momentum [see blue curve in Fig.~\ref{fig4}(b)] is still lower than
that from the SFA theory. This indicates that the electron
accelerates significantly due to the interplay of the Coulomb
potential and the laser field as it passes near the core. We
conclude that the amplitude difference between the different types
of trajectories is generally consistent with the phenomenological
picture of the ADK theory.

Physically, the above-mentioned behavior can be attributed to the fact that the
Coulomb potential decelerates the electron for Orbit I, which hinders
ionization. In contrast, for orbits II and III, the Coulomb potential accelerates the
electron. Hence, the electron acquires an additional pull, and may
escape moving along laser-dressed Kepler hyperbolae.

\subsubsection{Positions of interference maxima}

\begin{figure}[tb]
\begin{center}
\includegraphics[scale=0.3]{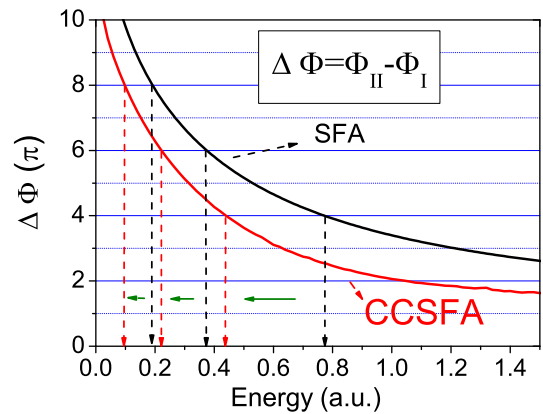}
\end{center}
\caption{(Color online) Phase difference $\Delta \Phi =\Phi_I-\Phi_{II}$ (in units of $\pi$) between trajectories of  type I and II as a
function of the photoelectron energy. The arrows denote the
positions of the interference maxima, where $\Delta \Phi=2k\pi$,
 $k=1,2,...$. The laser-field and atomic  parameters are given in Fig.~\ref{fig1}.}\label{fig5}
\end{figure}

We turn to the positions of interference maxima in the spectra.
Fig.~\ref{fig2} shows that the positions of the interference maxima
in the spectra from  the CCSFA theory are  shifted  when compared
with the SFA. Since the positions of interference maxima are
determined by the phase difference between different types of
trajectories we study how this is affected by the Coulomb potential. In Fig.~\ref{fig5},  we show the phase difference
between the type I and type II as a function of the photoelectron
energy. After considering the Coulomb potential, the dynamical phase
difference from the CCSFA becomes smaller than that from the SFA. This can be traced back to the fact that the type II trajectory accumulates a larger negative phase contribution from the Coulomb potential as it passes by the core. Overall, this reduces the phase difference to trajectory I. A similar analysis has been employed in our previous publication \cite{Lai2013PRA}, in the context of ATI with elliptically polarized fields.

Note that the reduction of the phase difference approaches $2\pi$,
so that neglecting multiples of $2\pi$ one can also interpret the
large shift towards lower energies as a small shift towards larger
energies. This ambiguity is resolved when we consider the effect of
the continuously rescaled Coulomb potential in Fig.~\ref{fig6},
which shows how a large shift towards smaller energies is
established as $C$ increases. These results are also consistent with
the recent TDSE simulations in \cite{arbo2010PRA}, and with the
outcome of similar Coulomb corrected approaches
\cite{Popruzhenko2008JMO,Yan2012PRA}.

\subsubsection{Overall ionization amplitude}

\begin{figure}[tb]
\begin{center}
\includegraphics[scale=0.3]{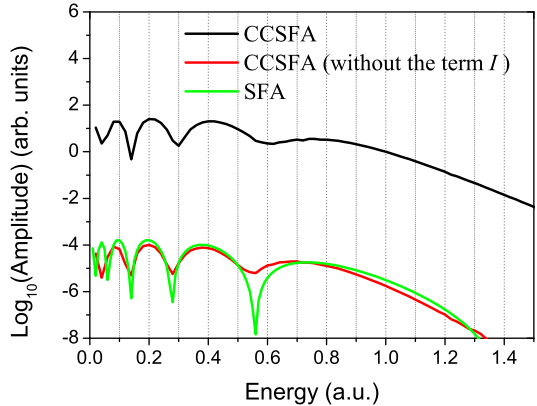}
\end{center}
\caption{(Color online) Photoelectron spectra from the CCSFA theory and the SFA theory,
respectively. The red curve is the spectrum from the CCSFA, but without
the subbarrier Coulomb term $I=-\int_{t_s}^{t_s^R} V[\textbf{r}(\tau)]d\tau$. The laser-field and atomic  parameters are given in Fig.~\ref{fig1}.}\label{fig7}
\end{figure}

Finally, we turn to the overall ionization amplitude of the spectra.
As we can see from Fig.~\ref{fig2}, the ionization amplitude from
the Coulomb-corrected theory is much larger than that from the SFA,
and much more in line with the TDSE result. Enhanced tunnel
ionization is a well-known effect caused by Coulomb corrections to
the effective potential barrier. It was first predicted in
\cite{Popov1967} and subsequently observed in several
Coulomb-corrected strong-field calculations
\cite{Smirnova2006JPB,Popruzhenko2008PRL}. Furthermore, in the 1980s
the Coulomb-induced orders-in-magnitude enhancement of tunnel
ionization rates of atoms and positive ions was well documented in
experiments \cite{Chin1988PRL}. Intuitively, it can be understood
that compared to the SFA theory, the barrier for a Coulomb potential
is smoother and lower, making it easier for the electron to tunnel
through. Within the CCSFA, this enhancement of the ionization
amplitude mainly originates from Coulomb term $I=-\int_{t_s}^{t_s^R}
V[\textbf{r}(\tau)]d\tau$  in the sub-barrier action
(\ref{in_barrier}). Fig.~\ref{fig7} shows the result if this term is
neglected. The overall magnitude of the spectrum is then comparable
to the SFA. In the CCSFA this term contributes with a negative
imaginary part $\mathrm{Im}\,I<0$, and thus increases the ionization
amplitude.

\section{Conclusions}
\label{conclusions}

In this work, we develop and use a path-integral formulation to
assess the influence of the residual Coulomb potential in
above-threshold ionization. We focus on the direct transition
amplitude, in which hard collisions with the core are not
incorporated. Overall, the photoelectron spectra obtained with the
Coulomb-corrected method presented in this paper exhibits an
excellent agreement with the \emph{ab initio} solution of the time
dependent Schr\"odinger equation, which is far superior to that
encountered for the plain strong-field approximation (SFA). This is
especially true for the interference substructure in the spectra.
We also perform a systematic analysis of how the Coulomb potential
modifies the orbits along which the electron may leave its parent
ion and reach the detector. We compare our results with those of the
SFA, and make an assessment of how, in the limit of vanishing
Coulomb coupling, the SFA is recovered. The present formulation is
closely related to the concept of quantum orbits widely employed in
semi-analytical strong-field approaches with and without Coulomb
corrections.

We have built upon the existing knowledge that the Coulomb potential introduces a richer topology in the electron motion \cite{Yan2010PRL}, with four distinct sets of orbits, and have related these orbits to those in the SFA in a more systematic way. Throughout, we have employed the same classification as in \cite{Yan2010PRL}, which specify these four sets as Orbits type I to IV. In particular, we have found that, for  electron emission along the polarization axis, due to the rotational symmetry with regard to the field axis, Orbits II and III will be located on a torus. This torus will contract for decreasing Coulomb coupling, until it becomes a point. Physically, this means that, for the SFA, Orbits type II and III will merge into a single, degenerate orbit if the final electron momentum is parallel to the laser-field polarization. For non-vanishing emission angle, the above-mentioned torus will break down, and there will be two discrete solutions. This behavior indicates that, strictly speaking, Orbits type II and III should not be treated independently in an asymptotic expansion when computing photoelectron spectra and momentum distributions. Indeed, a rigorous treatment would require solving the integral around the manifold exactly for a final momentum along the axis, and a uniform approximation for non-vanishing emission angle \cite{Tomsovic,Ullmo}. This strongly suggests that the cusps observed in \cite{Yan2010PRL,Yan2012PRA} close to the so-called ATI low-energy structure are related to this effect. It is indeed noteworthy that a very good agreement between the TDSE and the Coulomb-corrected SFA in \cite{Yan2010PRL,Yan2012PRA} was obtained throughout, except in this region (see also the discussion of this cusp in the review \cite{Popruzhenko2014JPB}). We have however not studied the above-mentioned cusp systematically.

Furthermore, our results indicate that, if the Coulomb potential is accounted for, the concepts of ``direct" and ``rescattered" electrons are not very clear-cut. These concepts are very clear in the SFA, as there are either hard collisions with the core, or no collisions at all. If the Coulomb corrections are present, however, the Coulomb potential strongly deflects Orbits type IV. These orbits go around the core, and there is a marked decrease in the electron's shortest distance from the origin as the photoelectron momentum increases. For high enough momentum, this distance is located in a region in which the binding potential is dominant. This behavior could be interpreted as a type of recollision, which is absent in the SFA. The amplitude associated with this type of trajectories is however very small and hence not relevant to the computation of ATI spectra in the parameter range of interest.

In addition to that, we have investigated the influence of the Coulomb potential on the ATI spectra, with emphasis on the interference contrast and position of the maxima. This influence has been traced back to particular sets of orbits. First, the contrast in the interference structure decreases, in comparison with the SFA. This happens because, in the SFA, Orbits I and II are equivalent, and displaced by half a cycle, while, if the Coulomb potential is included, this no longer holds. In fact, the Coulomb potential will decelerate the electron if it reaches the continuum along Orbit I, and will accelerate the electron if it is ionized along Orbits II and III.  This will lead to an increase in the amplitudes associated with Orbits II and III, and to a decrease in the amplitude related to Orbit I. Furthermore, there is the joint effect of Orbits II and III, which will weaken the fringes. Recently, the influence of Orbit III on interference effects has also been investigated in a different context, namely side-lobes in ATI electron momentum distributions, and it has been found to be significant \cite{Yan2013}.

The suppression of Orbit I and the enhancement of Orbits II and III has been confirmed by a systematic analysis of the initial momenta and ionization times. For Orbit I, the initial momentum increases when the Coulomb potential is considered, while, for type II/III orbits, the initial momentum decreases. Physically, this means that the Coulomb potential hinders ionization along Orbit I, as the electron will require a larger momentum to escape. For Orbits II/III, the Coulomb potential accelerates the electron after the tunneling ionization, so that a lower escape momentum is required. An increase in the initial momentum for the type I orbits also implies that the electron ionization time has moved away from the field maximum towards the field crossing. This means that the effective potential barrier through which it must tunnel will widen. Hence, there is also an increase in Im$[t_s]$. In contrast, for Orbits type II/III, the tunneling time moves to the crest of the laser field and thus the effective potential barrier becomes narrower. These observations are consistent with the changes in the real parts  Re$[t_s]$ of these times, as shown in Fig.~\ref{fig}.

Similarly to the results reported in \cite{Popruzhenko2008JMO,Yan2012PRA,Torlina2013PRA}, we also find that there is a phase shift towards lower energies in the interference maxima. In our model, this phase difference occurs due to Coulomb effects in the continuum propagation, while sub-barrier corrections mainly influence the overall yield.  In contrast, in \cite{Popruzhenko2008JMO,Yan2012PRA}, this phase difference is attributed to sub-barrier corrections instead. While the contour taken by us and the assumption that all variables are real outside the barrier are also employed in \cite{Popruzhenko2008JMO,Yan2012PRA}, the term $\dot{\textbf{p}}_s(\tau)\cdot
\textbf{r}_s(\tau)$ is absent in their action. Eq.~(\ref{saddle1}) shows that this term is proportional to the gradient of the binding potential. Its value is small for Orbit I, which moves towards the detector directly, while it is large for Orbits II and III, which are deflected by the core before reaching the detector. We have indeed verified that the phase from this term plays an important role in our formulation. Indeed, if this term is removed from the action, there is significant deviation between the TDSE and CCSFA results.
The stability factors employed here are also different from those in \cite{Popruzhenko2008JMO,Yan2010PRL,Yan2012PRA}, but they influence mainly the contrast and not the position of the maxima. They are, however, very important for a quantitative agreement with the full TDSE computations. This term is also absent in \cite{Torlina2013PRA}, in which the Eikonal-Volkov approximation is employed and the phase differences are obtained along propagation by using a complex intermediate coordinate. Since, however, in \cite{Torlina2013PRA} circularly polarized light is used, it is expected to be vanishingly small as the electron never returns to the core after tunneling ionization.
We expect that the present analysis will contribute to a better understanding of cusps and the ATI low-energy structure in the future.

\section*{Acknowledgments}
We are grateful to D. Bauer, T. M. Yan, S. Popruzhenko, O. Smirnova,
L. Torlina, M. Ivanov, T. Shaaran, J. Biegert, M. Lewenstein and D.
B. Milo\v{s}evi\'{c} for useful discussions. This work was funded by
the UK EPSRC (grants EP/J019240/1 and EP/J019585/1), the NNSF of
China (Grants Nos. 11204356 and 11474321) and by the Chinese Academy
of Sciences overseas study and training program. The data created
during this research is openly available \cite{data}. C.F.M.F. would
like to thank the Max Born Institute, Berlin and the Chinese Academy
of Sciences, Wuhan, for their kind hospitality.

\end{document}